\documentclass[12pt,preprint]{aastex}
\def\({\left(}
\def\){\right)}
\def\[{\left[}
\def\]{\right]}
\def\msun{{\rm ~M}_{\odot}}

\def\ledd{{L_{\rm Edd}}}
\usepackage{graphicx}
\usepackage{natbib}
\usepackage[usenames,dvips]{color}

\begin{document}
\title{ON THE PROPERTIES OF INNER COOL DISKS IN THE HARD STATE OF BLACK HOLE 
X-RAY TRANSIENT SYSTEMS}

\author{Ronald E. Taam}
\affil{Northwestern University, Department of Physics and Astronomy,
  2131 Tech Drive, Evanston, IL 60208; ASIAA/TIARA, P.O. Box 23-141, Taipei, 
10617 Taiwan; ASIAA/National Tsing Hua University - TIARA, Hsinchu, Taiwan}
\email{r-taam@northwestern.edu}

\author{B. F. Liu}
\affil{National Astronomical Observatories/Yunnan Observatory, Chinese
Academy of Sciences, P.O. Box 110, Kunming 650011, China}
\email{bfliu@ynao.ac.cn}

\author{F. Meyer}
\affil{Max-Planck-Institut f\"ur Astrophysik, Karl Schwarzschildstr. 1,
D-85740, Garching, Germany}
\email{frm@mpa-garching.mpg.de}

\and

\author{E. Meyer-Hofmeister}
\affil{Max-Planck-Institut f\"ur Astrophysik, Karl Schwarzschildstr. 1,
D-85740, Garching, Germany}
\email{emm@mpa-garching.mpg.de}
\begin{abstract}
The formation of a cool disk in the innermost regions of black hole X-ray transient systems in
the low hard state is investigated.  Taking into account the combined cooling associated with 
the Compton and conductive energy transport processes in a corona, the radial structure of a 
disk is described for a range of mass accretion rates.  The mass flow in an optically thick 
inner region can be  maintained by the condensation of matter from a corona with the disk 
temperature and luminosity varying continuously as a function of the accretion rate.  Although such a 
disk component can be present, the contribution of the optically thick disk component to the 
total luminosity can be small since the mass flow due to condensation in the optically thick 
disk underlying the corona can be significantly less than the mass flow rate in the corona. The 
model is applied to the observations of the low quiescent state of the black hole source GX 339-4 
at luminosities of $\sim 0.01 L_{Edd}$ and is able to explain the temperature of the thermal component 
at the observed luminosities.  Since conductive cooling dominates Compton cooling at low mass 
accretion rates, the luminosity corresponding to the critical mass accretion rate above which a weak 
thermal disk component can be present in the low hard state is estimated to be as low as $0.001 L_{Edd}$.
\end{abstract}


\section{Introduction}

The class of objects known as transient X-ray binary systems containing black holes is one of 
the few examples of astrophysical laboratories available for investigating the nature of accretion 
flows and the structure of accretion disks surrounding compact objects.  Their luminosities provide 
us with an opportunity to investigate a given system over a wide range of mass 
accretion rates onto the central black hole. Indeed, these systems have been observed in distinct 
states generally categorized as low/hard and high/soft states (see McClintock \& Remillard 2006).  In this 
categorization, the high/soft state is characterized by a strong thermal component 
corresponding to a color temperature of order 1 keV.  These states are observed during the outburst
phase in which the luminosity can approach the Eddington luminosity, $L_{\rm Edd} \sim 10^{39}$ ergs 
s$^{-1}$ for a $10 \msun$ black hole. On the other hand, the low/hard state is typically 
described by a power law with a photon power law index $\sim 1.7$.  This state is observed during the 
quiescent stages, where the luminosities can be as low as $10^{-7} L_{\rm Edd}$ (Corbel et al. 
2006). Hence, it is generally accepted that these states are distinguished by 
the spectrum and a range of luminosity with a transition at $\sim1-4\%$ of 
the Eddington luminosity (Maccarone 2003) from the low/hard state to the high/soft state.  

Given these observational diagnostic clues, progress in our theoretical understanding has advanced in 
parallel. Specifically, the spectrally soft state is believed to represent the emission from a cool optically 
thick and geometrically thin accretion disk extending to the innermost stable circular orbit ($3 R_S$ 
for a non-rotating black hole, where $R_S = 2GM/c^2$ is the Schwarzshild radius) as described in the 
seminal study by Shakura \& Sunyaev (1973, hereafter SS).  In contrast, the spectrally hard state is 
thought to arise from a geometrically thick, hot optically thin coronal or advectively dominated 
accretion flow (ADAF; Narayan \& Yi 1994; 1995a, b) located in the innermost regions of the disk near 
the black hole.  Here, the SS accretion disk is truncated at a distance of several hundred Schwarzschild 
radii from the black hole, leading to a transition from a disk dominated by a geometrically thin outer 
structure to a geometrically thick inner structure Esin et al. (1997, 1998, 2001).  Although 
a complete description of the truncation 
of the optically thick disk has yet to be worked out in detail, the mechanisms suggested to facilitate 
this transition have been investigated in the studies by Honma (1996) and Manmoto \& Kato (2000) 
based on a radial conductive energy transport process and by Meyer et al. (2000), Spruit \& Deufel 
(2002), and Dullemond \& Spruit (2005) based on a vertical evaporative process.
 
Recently, additional observational clues concerning the nature of the accretion flow and geometry 
in the low quiescent state have been found pointing to the possible existence of an inner cool 
optically thick disk structure in black hole transient X-ray binary systems.  Although the 
observational signatures are difficult to extract because of the low X-ray count rates, the evidence 
for such a structure has been inferred from the source GX 339-4 (see Miller et al. 2006a, b; Tomsick 
et al. 2008) where a weak, soft, thermal emission component ($\lesssim 0.3$ keV) has been detected 
and blurred reflection and iron features are indicated in fits to the X-ray spectrum for luminosities 
as low as $\sim 0.01 L_{\rm Edd}$. One of the goals of current observational studies has been to 
extend such studies to other sources (e.g., J 1817-330, see Rykoff et al. 2007; and J1753.5-0127, 
see Miller et al. 2006a; Ramadevi \& Seetha 2007; Soleri et al. 2008) and to determine the 
luminosity range for which such a thermal component exists so that a full description of the 
accretion flow geometry in the various states of transient X-ray binary systems can be constructed. 

Given the existence of this weak thermal component, the accretion disk structure in the quiescent state 
would be described by three distinct regions.  Specifically, a small inner cool, optically thick disk, 
a geometrically thick hot coronal region at intermediate radial scales, and a very extensive geometrically 
thin disk nearly filling the Roche lobe of the black hole would contribute to the weak thermal X-ray 
component, the power law X-ray component, and the infrared and optical emission respectively. Such a 
configuration was already contemplated in early work by R\`o\.za\`nska \& Czerny (2000) and more 
recently by Liu et al. (2006), Meyer et al. (2007) and Mayer \& Pringle (2007). To maintain such a remnant 
cool disk in the innermost regions (in the presence of mass transport by turbulent viscous stresses) 
gas must condense from an overlying corona. 

In this study we focus on the general description of the structure and properties of the disk in its 
innermost regions during the quiescent state of a black hole X-ray transient binary system, taking into 
account the effects of both thermal bremsstrahlung and comptonization.  This study differs from our recent 
study (Liu et al. 2007) where only one such process was considered dominant for any given structure. In 
\S 2, we describe the theoretical model and the modifications required in considering
more than one cooling process in any given region of the disk.  The solutions to the modified set of 
equations are presented for a range of mass accretion rates and viscosity parameters in \S 3.  
The numerical results are compared to the observations of GX 339-4 in \S 4 and their implications 
are discussed in the last section. 

\section{Theoretical Model}

Early investigations (Meyer \& Meyer-Hofmeister 1994, Meyer, Liu, \& Meyer-Hofmeister 
2000; Liu et al. 2002) have shown that a hot corona overlying a cool disk leads to the
evaporation of cool gas into a corona, through which mass is continuously supplied for 
accretion onto a central compact object. The evaporation is driven by vertical thermal 
conduction, and the accretion of evaporated mass through the corona is a result of 
viscous transport. The properties of the evaporation fed corona are a function of the 
distance from the accreting compact object.  In the case of a black hole transient 
X-ray binary system in the quiescent state, the mass flow rate in the corona reaches a 
maximum at a distance of $\sim 300 R_S$ by efficient evaporation 
and decreases inward as a result of condensation of coronal gas to a cooler disk. 

When the mass supply rate is sufficiently low, the optically thick outer cool disk is 
truncated in the region where the integral of the evaporation rate over the radial extent 
is greater than the mass inflow rate, leading to the formation of a gap between outer and 
inner disk, filled by a coronal flow/ADAF. The radial extent of the coronal 
gap region is dependent on the mass supply rate to the outer disk with a larger coronal gap 
region corresponding to a lower mass accretion rate. For black hole X-ray binary systems in 
the quiescent state, the accretion rate can be very low, and the outer disk is sufficiently distant 
from the central black hole that it can be neglected in describing the innermost disk regions. 
In this latter region, the dominant accretion flow lies within the corona/ADAF, accompanied 
by a weak mass inflow within an underlying cool disk.  The condensation of coronal gas to the 
disk follows from energy (and pressure) balance between this inner cool disk and its overlying 
corona, providing the only source of mass in the inner optically thick disk.  As shown by Liu 
et al. (2007) such a configuration can be maintained for a range of accretion rates and, hence,
luminosities. 

In this work, we generalize the description of the inner disk and corona configuration by 
considering Compton scattering and conduction, contributing simultaneously, to the 
cooling of the disk. The treatment of both cooling processes leads to a modification of 
the temperature of the electrons in the corona, and, hence, the condensation rate. In addition, 
we also include contributions from both inverse Compton emission from the corona and 
bremsstrahlung emission from the radiating/transition layer to the total luminosities. This 
treatment of the cooling processes is in contrast to our previous study (Liu et al. 2007) where 
always only one of the dominant emission mechanism was taken into account (but similar in our 
neglect of non-thermal cooling processes).  As a consequence of the new treatment, the 
physical characteristics of the inner regions (e.g., the effective temperature of the inner 
cool disk) vary continuously with variations in the luminosity.  To facilitate comparisons 
with Liu et al. (2007), the effect of X-ray irradiation of the hot coronal gas on the cool disk 
is neglected; however, an approximate estimate of its effect is given and its implications discussed 
in \S5.

\subsection{Condensation caused by either conductive cooling or
Compton cooling}

The energy balance in the transition layer between the disk and corona depends
on the ADAF lying above the disk (Narayan \& Yi 1995b). As derived in our 
earlier work (Meyer et al. 2007; Liu et al. 2007)
 the condensation rate (mass flow rate in vertical direction)
per unit area is determined as 

  \begin{equation}\label{cnd-general}
\dot m_z={\gamma-1\over \gamma} \beta{-F_{\rm c}^{\rm ADAF}\over {\Re T_{i}\over 
\mu_i}}\(\sqrt{C}-1\),
\end{equation}
where
\begin{eqnarray}\label{C}
C \equiv\kappa{_0} b
\left(\frac{0.25\beta^2 p_0^2}{k^2}\right)
\left(\frac{T_{\rm {cpl}}}{F_c^{\rm{ADAF}}}\right)^2 \nonumber
\end{eqnarray}
 with $F_{\rm c}^{\rm ADAF}$ the conductive flux from the upper ADAF
arriving 
at the interface of the transition layer, $T_{\rm cpl}$ the coupling
temperature and 
$p_0=(2/\sqrt \pi)p$ the pressure in the transition layer,
\begin{eqnarray}\label{scaled}
&F_{\rm c}^{\rm ADAF}= -(\kappa_0 K n_i n_e T_i)^{1/2}T_{\rm em},\nonumber\\
&T_{\rm {cpl}}=1.98\times 10^9 \alpha^{-4/3}\dot m^{2/3}\rm K.
\end{eqnarray}
Pressure, ion temperature and electron number density, $p$, $T_i$ and $n_e$ 
are quantities of the ADAF, 
\begin{eqnarray}\label{scaledadd}
&T_i+1.077 T_e =1.98\times10^{12} r^{-1}\rm K,\\\nonumber
&p=1.87\times 10^{16}\alpha^{-1}m^{-1}\dot m r^{-5/2}
\, \rm{g cm^{-1} s^{-2}},\\
&n_e=5.91\times10^{19}\alpha^{-1}m^{-1}\dot m r^{-3/2} \,  \rm{cm^{-3}} \nonumber .  
\end{eqnarray}
The meaning of the further quantities in the equations is: 
$\beta$ is the ratio of gas 
pressure to total pressure, $\gamma$ the ratio of specific heats, $\gamma=(8-3\beta)/(6-3\beta)$,
 $\mu_i$ the molecular weight of ions 
(1.23 for the assumed standard chemical abundance) and $k$ is the
Boltzmann constant. 
$\kappa{_0} =10^{-6}{\rm erg\,s^{-1}cm^{-1}K^{-7/2}}$ is the 
thermal conductivity 
coefficient in the relation for the conductive flux, 
$F_{\rm c}=-\kappa_0 T_e^{5/2}dT_e/dz$ (Spitzer 1962). The coefficient 
$b=10^{-26.56}\rm g\,cm^5\,s^{-3}\,K^{-1/2}$ is used for the description 
of the free-free radiation (Sutherland \& Dopita 1993). The energy transfer 
rate from ions to electrons is taken as $q_{\rm ie}=Kn_in_eT_iT_e^{-3/2}$ 
with $K= 1.64\times 10^{-17} \rm {g cm^{5}s^{-3}deg^{1/2}}$ (Stepney
1983) and $n_i=n_e/1.077$ is the ion number density. Following 
convention, we take $\alpha$ as the viscosity coefficient. The quantities $m$, 
$\dot m$, and $r$ are expressed in units of a solar mass, the Eddington rate, 
$\dot M_{\rm Edd}=L_{\rm Edd}/0.1c^2$, and the Schwarzschild 
radius, $R_S$, respectively. $T_{\rm em}$ is the maximal electron temperature in 
the corona, to be determined for either conduction or Compton cooling,
as derived in the earlier work (Liu et al. 2007).

Once $T_{\rm em}$ is known the condensation rate 
can be determined by Eqs.(\ref{cnd-general})--(\ref{scaledadd}) at any 
radius for a given black hole mass, mass accretion rate and
 viscosity.

In the case of inefficient Compton cooling in the corona, 
vertical conduction is the dominant cooling mechanism.
Here, by collisional heat input from the ions (${dF_c/dz}=q_{\rm ie}$),
a vertical temperature gradient forms which results in a maximal 
temperature at the upper boundary of the ADAF and a conductive flux at the 
lower boundary,

\begin{eqnarray}\label{T_em}
&T_{\rm em}=2.01\times 10^{10}\alpha^{-2/5}\dot m^{2/5} r^{-2/5}\rm K,\\
&F_{\rm c}^{\rm ADAF}=-6.52\times 10^{24} \alpha^{-7/5}m^{-1}\dot m^{7/5} r^{-12/5}\nonumber.
\end{eqnarray}
On the other hand, for a sufficiently high radiation flux from the underlying disk, the disk 
radiation propagating through the corona can lead to efficient Compton cooling of the corona. The 
electrons are cooled by Compton scattering to a temperature (determined by $q_{\rm ie}=
q_{\rm cmp}$) lower than the value given by conductive cooling (Eq.(\ref{T_em})). As a result, 
the heat flux at the transition layer decreases and the condensation rate increases, as 
determined by the energy balance in the transition layer between cooling by bremsstrahlung 
emission and heating by the conductive flux from the upper ADAF and by the enthalpy flux of the 
condensing gas. The maximal temperature in the corona and the corresponding conductive flux to 
the transition/radiating layer are expressed as
\begin{eqnarray}\label{T_ec}
& T_{\rm em}=3.025\times 10^9\alpha^{-2/5} m^{-2/5}\dot m^{2/5}r^{1/5}\[1-\({3\over r}\)^{1/2}\]^{-2/5}\({T_{\rm eff,max}\over 0.3keV}\)^{-8/5}\rm K,\\
& F_c^{\rm ADAF}=-9.816\times 10^{23}\alpha^{-7/5} m^{-7/5}\dot m^{7/5}r^{-9/5}
\[1-\({3\over r}\)^{1/2}\]^{-2/5}
\({T_{\rm eff,max}\over 0.3keV}\)^{-8/5}
\rm erg\,s^{-1}\,cm^{-2}.\nonumber
\end{eqnarray}
The effective temperature of the disk, $T_{\rm eff}$, corresponds to the
mass flow in the disk accumulated by the condensation of matter from the corona.
$T_{\rm eff}$ reaches its maximum value $ T_{\rm eff, max}=1.335\times 
10^7m^{-1/4}\dot m_{\rm cnd}^{1/4}$ at a distance  
$r_{\rm tmax}=3\,\frac{49}{36} R_s$ (for details see Liu et al. 2007).

\subsection{Condensation caused by the two processes: conductive cooling and
Compton cooling operating simultaneously}

The above describes the formulation of our model as outlined in a previous study (Liu et al. 
2007) when one of the cooling processes (either thermal conduction or Compton scattering) 
is dominant.  However, in general, both processes can contribute to
the cooling simultaneously. In particular, 
for a certain range of mass accretion rates, the rate of cooling associated with the 
conduction and Compton processes are comparable.  In this case, neither the Compton dominated 
model nor the conduction dominated model provides an adequate description for the disk/corona 
configuration.  In this case, both mechanisms must be treated together. However, 
construction of the detailed vertical structure for the condensation region cannot be expressed 
in analytical form. In the spirit of the analysis of Liu et
al. (2007), we modify the formulae describing the 
conduction dominant and Compton dominant cases in such a way that we
obtain a smooth transition when both mechanisms are of comparable
importance. Our approach is described below.

For given values of $\alpha$, $m$, and $\dot m$ at a given distance, $r$, the importance of 
Compton or conduction cooling is determined by comparing the electron temperature given by 
considering the operation of Compton cooling alone (i.e. $q_{\rm ie}=q_{\rm cmp}$) and 
conductive cooling alone (i.e. $q_{\rm ie}=dF_c/dz$). The process that results in a lower 
temperature is the more efficient cooling mechanism and, thus, is taken as the dominant one. 
To determine the strength of the neglected cooling process compared to that of the dominant process, 
we define the ratio, $\lambda$, between the Compton cooling rate, $q_{\rm cmp}$, and the 
conduction cooling rate, $dF_c/dz$,  
\begin{equation}\label{lambda-d}
\lambda\equiv {q_{\rm cmp}\over dF_c/dz}.
\end{equation}
The Compton cooling rate is given by 
\begin{equation}
q_{\rm cmp} = {4kT_e\over m_ec^2} n_e \sigma_T c {aT_{\rm eff}^4(r)\over 2}
=f_1(\alpha, m,\dot m) \dot m_{\rm cnd} T_e r^{-9/2}\[1-\(\frac{3}{r}\)^{1/2}\],
\end{equation}
where $a$ is radiation density constant, $\sigma=\frac{ac}{4}$, $\sigma_T$ Thomson cross section,
$\sigma T_{\rm eff}^4(r)=(3GM \dot M_{\rm cnd}/8\pi R^3) \[1-(3R_S/R)^{1/2}\]$, and $f_1$ is some 
function only depending on the ADAF parameters.

$dF_c/dz$ can 
be approximated by taking the value of the flux at the upper boundary ($F_c=0$) and at the 
lower boundary, $F_c^{\rm ADAF}=-(k_0Kn_in_eT_i)^{1/2}T_{\rm em}$, to give 
$$dF_c/dz\propto -F_c^{\rm ADAF}/H\propto -F_c^{\rm ADAF}r^{-1}.$$ 
Replacing $F_c^{\rm ADAF}$ by  Eqs.(\ref{scaled}) and (\ref{scaledadd}) we have 
\begin{equation}\label{Fc-apprx}
dF_c/dz= f_2\(\alpha, m,\dot m\)r^{-3} T_{\rm em}. 
\end{equation}
Thus, from  Eqs.(\ref{lambda-d})--(\ref{Fc-apprx}) and $T_e=T_{\rm em}$, $\lambda$ can be 
approximately expressed as
\begin{equation}\label{lambda0}
\lambda=f\(\alpha, m,\dot m\)\dot m_{\rm cnd} r^{-3/2}\[1-\(\frac{3}{r}\)^{1/2}\].
\end{equation}  
The function $f$ again only depends on the ADAF parameters. 
The value of $\lambda$ is very small when conduction is the overwhelmingly dominant cooling 
process, reaching unity when the Compton process is equally important, and is much  
larger than unity for cooling dominated by Compton scattering.  In the critical case when 
the cooling process switches from conduction dominant to Compton dominant, $\lambda_c=1$, 
the temperature inferred from the Compton cooling process and the conduction cooling process
is the same, which gives (by Eq.(\ref{T_em})=Eq.(\ref{T_ec})),  
\begin{equation}
\dot m_{\rm cnd}r_{\rm c}^{-3/2}\[{1-\({3\over r_{\rm c}}\)^{1/2}}\]={1\over 2.464\times 10^4}.
\end{equation}
This determines $f\(\alpha, m,\dot m\)=2.464\times 10^4$ by the conditions defining the critical 
case from Eq.(\ref{lambda0}). Thus, $\lambda$ can be written, in general, as 
\begin{equation}\label{lambda}
\lambda=2.464\times 10^4\dot m_{\rm cnd}r^{-3/2}\[1-\(\frac{3}{r}\)^{1/2}\]
\end{equation}
and used in calculating the condensation rate. To obtain a solution that includes both effects, 
thermal conduction and Compton cooling,  we modify the energy equation, leading to a change in the 
electron temperature and, hence, a change in the condensation 
rate and luminosities from both the corona and the disk. In the case of conduction dominant 
cooling, the energy equation is modified as
\begin{equation}
q_{\rm ie}=(1+\lambda)dF_c/dz.
\end{equation}
Since $q_{\rm ie}\propto T_e^{-3/2}$ and $dF_c/dz\approx -F_c^{\rm ADAF}r^{-1}\propto T_{\rm em}$, 
the energy equation leads to a modified temperature 
\begin{equation}
T_{\rm em}=2.01\times 10^{10}\alpha^{-2/5}\dot m^{2/5} r^{-2/5}(1+\lambda)^{-2/5}\rm K.
\end{equation}
Similarly, in the case of Compton dominant cooling, the energy equation should be modified as
\begin{equation}
q_{\rm ie}=(1+{1\over\lambda})q_{\rm cmp}.
\end{equation}
Note that $q_{\rm cmp}\propto T_e$, and the modified energy equation results in a 
modification factor $(1+{1\over\lambda})^{-2/5}$ to the temperature,
\begin{equation}
T_{\rm em}=3.025\times 10^9\alpha^{-2/5} m^{-2/5}\dot m^{2/5}r^{1/5}\[1-\({3\over r}\)^{1/2}\]^{-2/5}(1+{1\over\lambda})^{-2/5} \({T_{\rm eff,max}\over 0.3keV}\)^{-8/5}\rm K.
\end{equation}
With the modified temperature, the heat flux and the resulting integrated condensation rate 
in a region between  $R_i$ and $R_o$ can be calculated for given $\alpha$, $m$ and $\dot m$,
\begin{equation}\nonumber
\dot m_{\rm cnd}=\int_{R_i}^{R_o} {4\pi R \over \dot M_{\rm Edd}}\dot m_zdR.
\end{equation}
The condensation rate for Compton dominant cooling is
\begin{equation}\label{condensation-cmp}
\dot m_{\rm cnd}=A\left\{2B\[\({r_{\rm o}\over r_i}\)^{1/2}-1\]-\int_{r_i/3}^{r_{\rm o}/3}(1+{1\over\lambda})^{-2/5}x^{1/5}\(1-x^{-1/2}\)^{-2/5}dx \right\},
\end{equation}
and for conduction dominant cooling 
\begin{equation}\label{condensation_cnd}
\dot m_{\rm cnd}= 3.23\times 10^{-3}\alpha^{-7}\dot m^3 f(x)
\end{equation}
with
\begin{equation}\label{functionx}
f(x)=\({r_o\over r_1}\)^{3/5}\[{6\({r_1\over r_o}\)^{1/10}-6{\({r_1\over r_o}\)^{1/10} x^{1/2}-3 \int_{r_i/r_o}^1 \(1+\lambda\)^{-2/5} x^{-2/5}}dx}\],
\end{equation}
where $r_{1}=0.815\alpha^{-28/3}\dot m^{8/3}$ is the condensation radius determined by $C=1$ in the case of conduction cooling  
before modification. The new factor $r_o\over r_1$ is a consequence of generalizing the 
condensation formula to reflect the condensation rate at any condensation region. When 
$\lambda=0$ and $r_o= r_1$, Eq.(\ref{functionx}) reverts to our old form. When $\lambda\rightarrow
\infty$, Eq.(\ref{condensation-cmp}) also reverts to the previous form for the case of Compton 
cooling.

Note that since $\lambda$ depends on the condensation rate, iteration is required until a 
a self-consistent condensation rate is obtained. In practice, the modification of the 
condensation rate is very small at low accretion rate since the conduction process dominates.
Similarly, the modification is also small if Compton cooling is very 
strong compared to conduction.  We note, however, that these conditions are not necessarily 
satisfied for black hole X-ray binary systems in their low hard or intermediate states.  
In this regime, the modification factors are important since conduction and Compton 
scattering processes can be of comparable strength in cooling the corona.  

\subsection{The radial extent of Compton and conduction dominated regions}
\begin{figure}
\plottwo{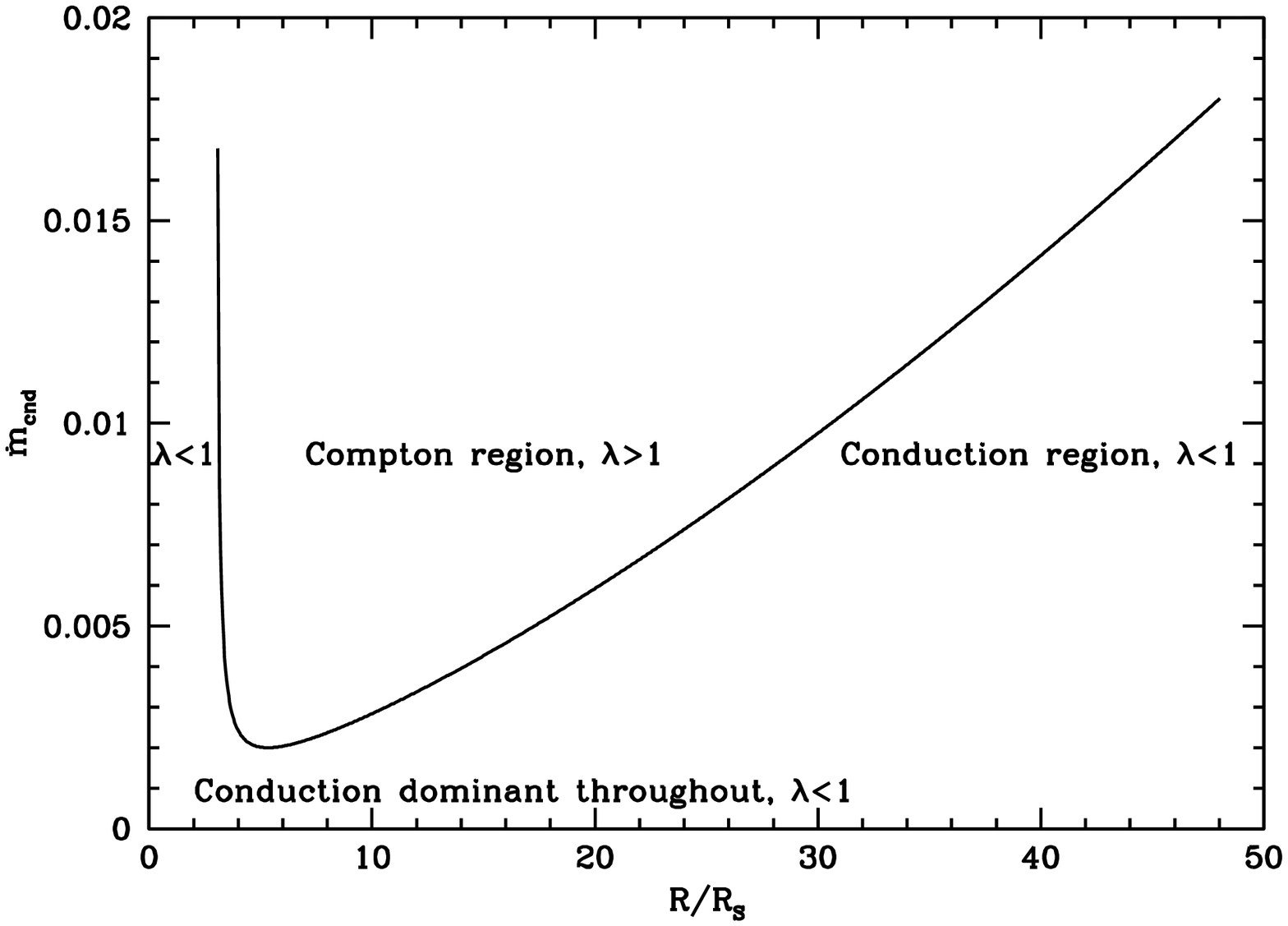}{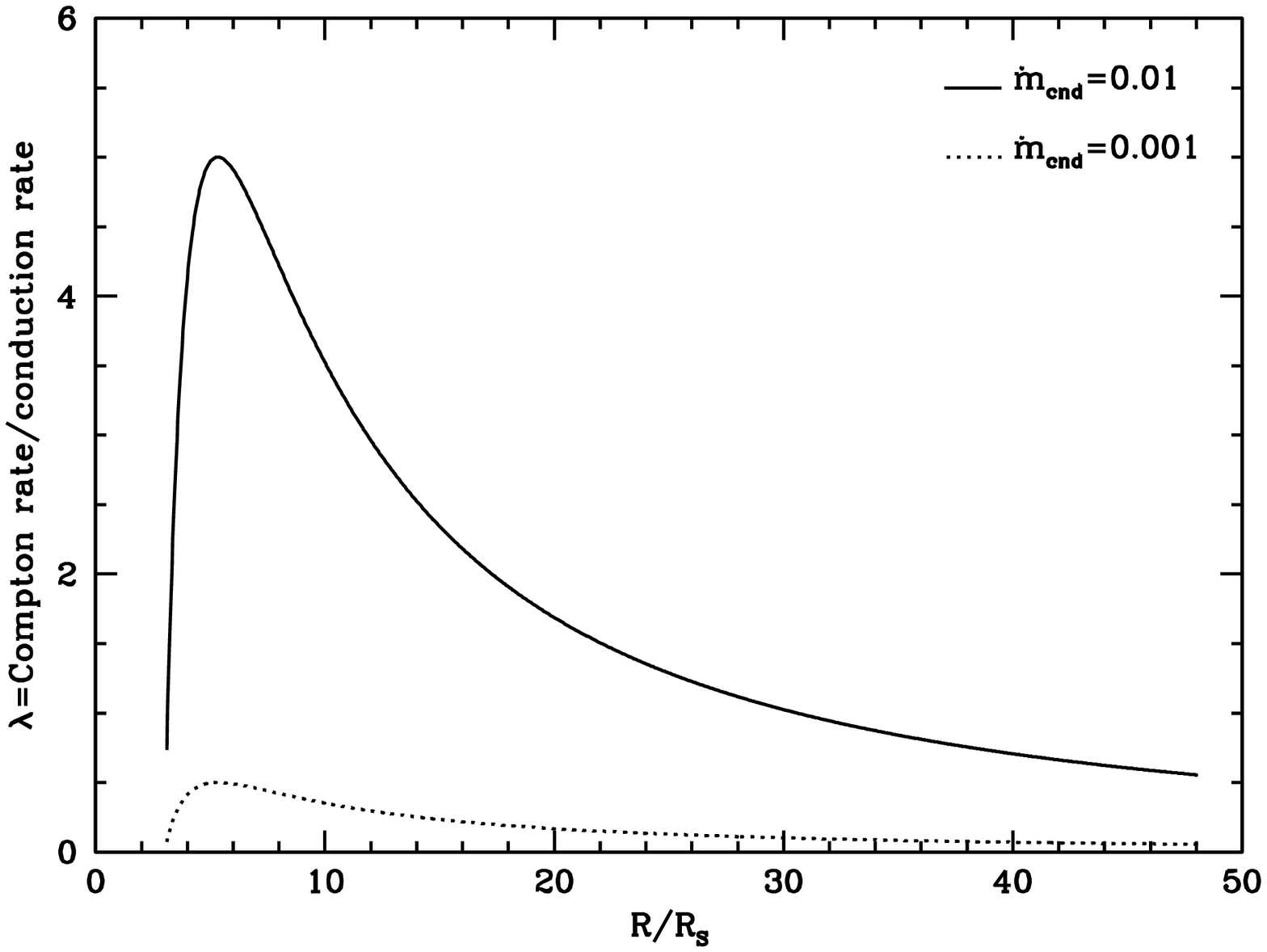}
\caption{\label{cmpregion}The left panel illustrates regions dominated by the Compton or conduction 
cooling process as a function of the total condensation rate. The Compton scattering is dominant 
at higher condensation rates (or higher accretion rates). The right panel shows the regions where Compton 
cooling is important and its relative strength, $\lambda$, as a function of distance for condensation 
rates of $\dot m_{\rm cnd}=0.001$ (dotted curve) and $\dot m_{\rm cnd}=0.01$ (solid curve). Between 
these two condensation rates, inclusion of both cooling mechanisms is necessary, especially at distances 
where $\lambda \sim 1$.} 
\end{figure}

In describing the thermal state of the corona above an inner disk, (i) cooling in the corona 
could be conduction dominant throughout the corona, (ii) Compton dominant throughout, or (iii) 
conduction dominant in the outer regions and Compton dominant in the inner region, depending 
on the binary system parameters. Therefore, for given values of $\alpha$, $m$ and $\dot m$, 
an estimate of the radial distance from the black hole for the Compton dominant region is 
obtained by assuming that the Compton process cools the electrons to the same temperature as the 
conduction process,
\begin{equation}\label{r_cmp}
r_{\rm cmp}\[{1-\({3\over r_{\rm cmp}}\)^{1/2}}\]^{-2/3}= 846.71\dot m_{\rm cnd}^{2/3},
\end{equation} 
The above equation either admits two solutions or no solution for $r_{\rm cmp}$. In 
case no solution exists, Compton cooling never dominates in the corona and the modified 
conduction model is an adequate description for the inner disk and corona. The existence of the 
two solutions $r_{\rm cmp1}$ and $r_{\rm cmp2}$ determines the Compton dominant region $r_{\rm 
cmp1}<r<r_{\rm cmp2}$. If $r_{\rm cmp2}$ is larger than the condensation radius calculated 
by the conduction model, which is modified as $r_{d}=0.815\alpha^{-28/3}\dot m^{8/3}
\[1+\lambda(r_d)\]^4$, the cooling is dominated by the Compton process at all 
radii, and we use the modified Compton model for the numerical calculations.  On the 
other hand, if  $r_{\rm cmp2}$ is smaller than $r_{d}$, the coronal region is divided into 
a conduction and Compton dominant region. In Figure \ref{cmpregion} the radial extent of 
Compton and conduction dominant regions is presented. The left panel delineates the regions 
dominated by Compton or conduction cooling as a function of the total condensation rate.  
At the lowest condensation rates, conduction is dominant throughout the disk.  On the other hand,  at higher 
condensation rates,
the Compton dominant region always lies interior to the conduction dominant region. The right panel reveals the relative strength of the Compton cooling rate 
as compared to conduction as a function of distance at given condensation rates. The modified 
condensation rate is calculated employing the relevant models and values of $\lambda$ in each 
region.  Their contributions are summed to determine the modified total condensation rate of the system.
 
The luminosity from bremsstrahlung radiation in the radiating/transition layer does not 
vary with the electron temperature in the corona since the density and coupling temperature 
are determined by $\alpha$, $m$, and $\dot m$, independent of the coronal temperature. Thus, 
it has the same form for both the Compton and the conduction dominant regions
and is unaffected by the modification when both cooling mechanisms are included. We integrate 
the bremsstrahlung emission throughout the disk-corona region yielding,
\begin{equation}\label{L_cnd}
{L_{\rm Brem}\over L_{\rm Edd}}=
0.0642 \alpha^{-7/3}\dot m^{5/3}\[1-\({3\over r_d}\)^{1/2}\].
\end{equation}
Since the Compton luminosity from the corona depends on the electron temperature, its 
determination must include the contributions from both cooling processes. In the Compton 
cooling dominant region, a correction factor $(1+{1\over\lambda})^{-2/5}$ is added to the 
integrand, 
\begin{equation}\label{L_cmp1}
{L_{\rm cmp}\over L_{\rm Edd}}=0.392\alpha^{-7/5}m^{3/5}\dot m^{7/5}\({T_{\rm eff, max}\over 0.3keV}\)^{12/5}\int_{r_{\rm i}/3}^{r_{\rm o}/3}(1+{1\over\lambda})^{-2/5}x^{-23/10}\(1-x^{-1/2}\)^{3/5}dx.
\end{equation}
In the conduction dominant region, the luminosity from Compton scattering was neglected in 
our previous work. To take this contribution into account,  the electron temperature is replaced by 
Eq.(\ref{T_em}) in the Compton cooling rate, leading to a Compton luminosity from 
the conduction dominant region from $r_i$ to $r_o$,
\begin{equation}\label{L_cmp2}
{L_{\rm cmp}\over L_{\rm Edd}}=1.349\alpha^{-7/5}m\dot m^{7/5}\({T_{\rm eff, max}\over 0.3keV}\)^4\int_{r_{\rm i}/3}^{r_{\rm o}/3}(1+\lambda)^{-2/5}x^{-29/10}\(1-x^{-1/2}\)dx.
\end{equation}
The total corona luminosity is, thus, the sum of bremsstrahlung radiation
 (Eq.(\ref{L_cnd})) in the radiation/transition layers and Compton scattering 
in the ADAF above
(Eqs. (\ref{L_cmp1}) and (\ref{L_cmp2})) when both regions exist.  

The disk luminosity is determined under the assumption of emission by a local blackbody and 
calculated as an integral over the disk as,
\begin{equation}\label{Ldisk}
{L_{d}\over L_{\rm Edd}}=\int_{3R_s}^{R_d}\sigma T_{\rm eff}^4(R){4\pi R\over L_{\rm Edd}} dR=\int_{3R_s}^{R_d}{3GM\dot M\[1-({3R_s\over R})^{1/2}\]\over 8 \pi R^3}{4\pi R\over L_{\rm Edd}} dR.
\end{equation}
Note that the mass flowing in the inner disk through any radius $R$ originates from the condensed gas 
integrated from the critical condensation radius ($R_d$) to $R$. Hence, the mass accretion rate in the disk is 
a function of radius, which increases with decreasing radii, representing the cumulative effect of 
the condensation rate of gas onto the disk. The disk luminosity is then 
calculated from all radii with the corresponding accretion rate profile.

In our procedure, the condensation rate is first calculated assuming conduction as the only cooling 
mechanism. If this condensation rate heats the disk to a temperature sufficiently  high that 
Compton cooling is more efficient than conduction in some regions, we adopt the Compton 
cooling model in the Compton dominant region but retain the conduction model in the conduction 
region. The condensation rate is recalculated from these different regions as iterative 
calculations are required for the Compton dominant region.  
These results are modified taking into account the neglected cooling by calculating the 
value of $\lambda$ from the inferred condensation rate.  As the modified condensation rate 
results in a new value of $\lambda$, the process is repeated until the presumed condensation 
rate approaches the derived condensation rate.  Finally, the contributions to the total luminosity 
from Compton scattering and bremsstrahlung throughout the coronal region and multi-colour blackbody radiation from the inner disk are 
calculated with the derived condensation rate and correction factor $\lambda$ by equations 
(\ref{L_cnd}), (\ref{L_cmp1}), (\ref{L_cmp2}) and (\ref{Ldisk}).

\section{Numerical Results}

To investigate the properties of the inner optically thick disk in X-ray transient binary systems, 
we adopt a  black hole mass of $10M_\odot$ and the low mass accretion rate range corresponding to 
the transition to an ADAF.  In the following, we describe the numerical results for the inner disk temperature, 
size, and luminosity from the inner disk as determined from the condensation rates from the corona 
as a function of the total mass accretion rate.  To determine their  sensitivity to the viscosity in 
the disk, we also examine their dependence on the viscosity parameter.  

\begin{figure}
\plotone{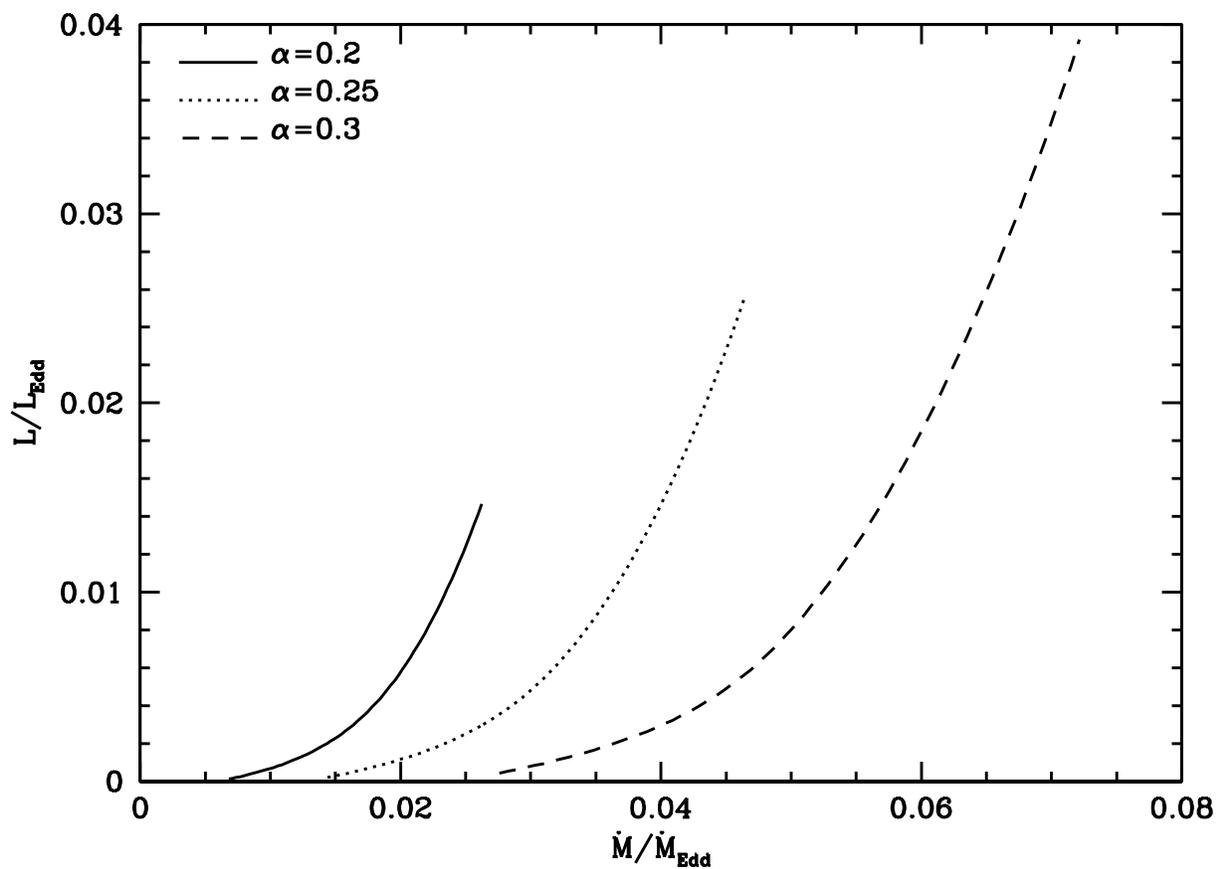}
\caption{\label{f:mdot-L}The luminosity as a function of the mass accretion rate. The total luminosity 
increases with accretion rate and the energy conversion efficiency in the accretion flow, as measured 
by ${\eta\over 0.1}={L\over L_{\rm Edd}}/{\dot  M\over \dot M_{\rm Edd}}$, is greater at higher accretion 
rates and lower values of the viscosity parameter $\alpha$.} 
\end{figure}

The luminosity from the disk and corona is illustrated as a function of accretion rate in Figure 
\ref{f:mdot-L}. 
It can be seen that the Eddington scaled luminosity does not scale linearly with the mass accretion 
rate, indicating that the conversion of gravitational energy to radiative energy by an ADAF+disk in  
low states is not constant nor as efficient ($\eta=0.10$) as in a SS accretion disk.  The deviation 
of the $L/L_{\rm Edd}-\dot M/\dot M_{\rm Edd}$ curve from a linear relation indicates that the energy 
conversion efficiency decreases with lower mass accretion rates, qualitatively consistent with an 
ADAF model. 

\begin{figure}
\plottwo{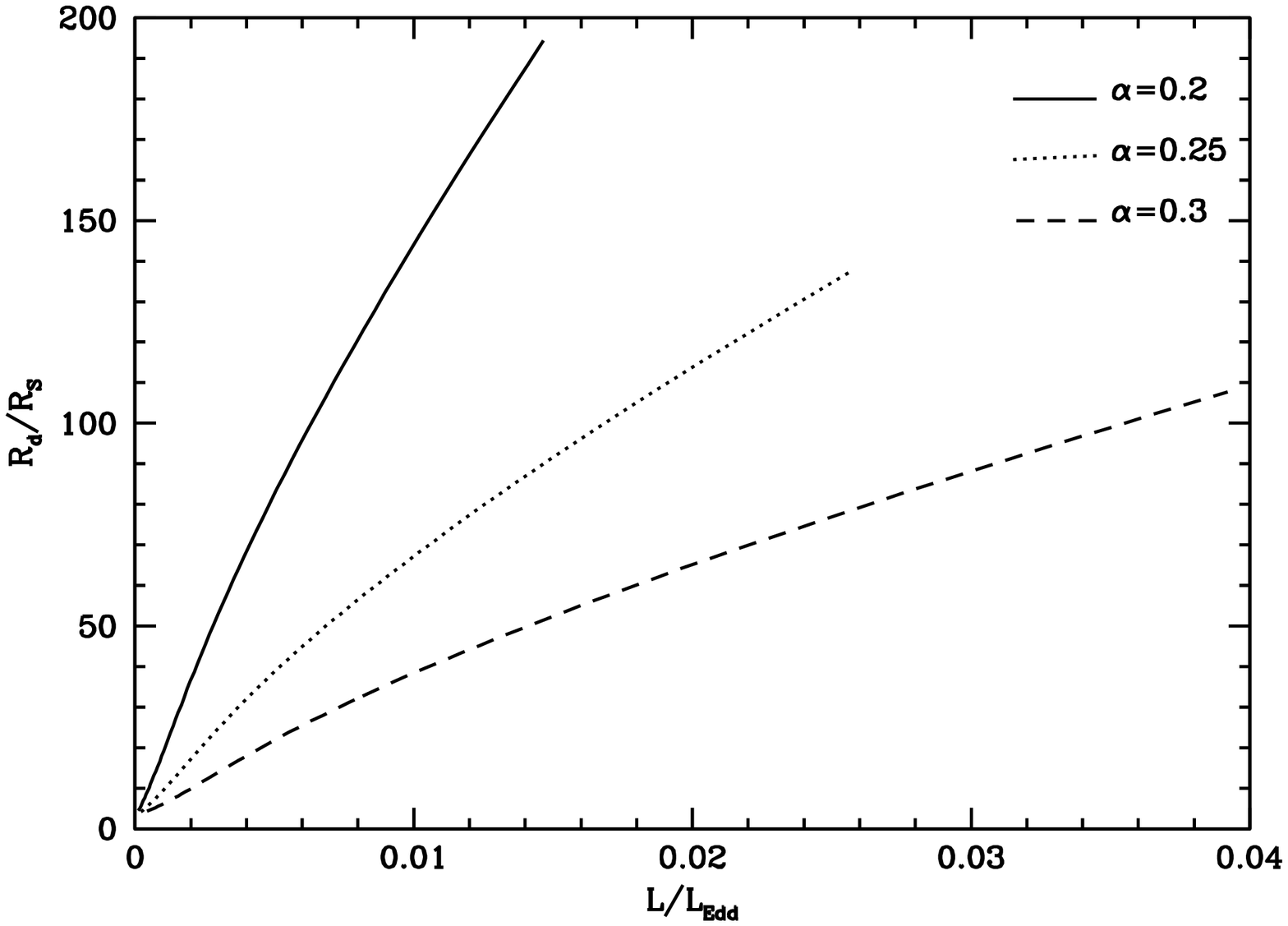}{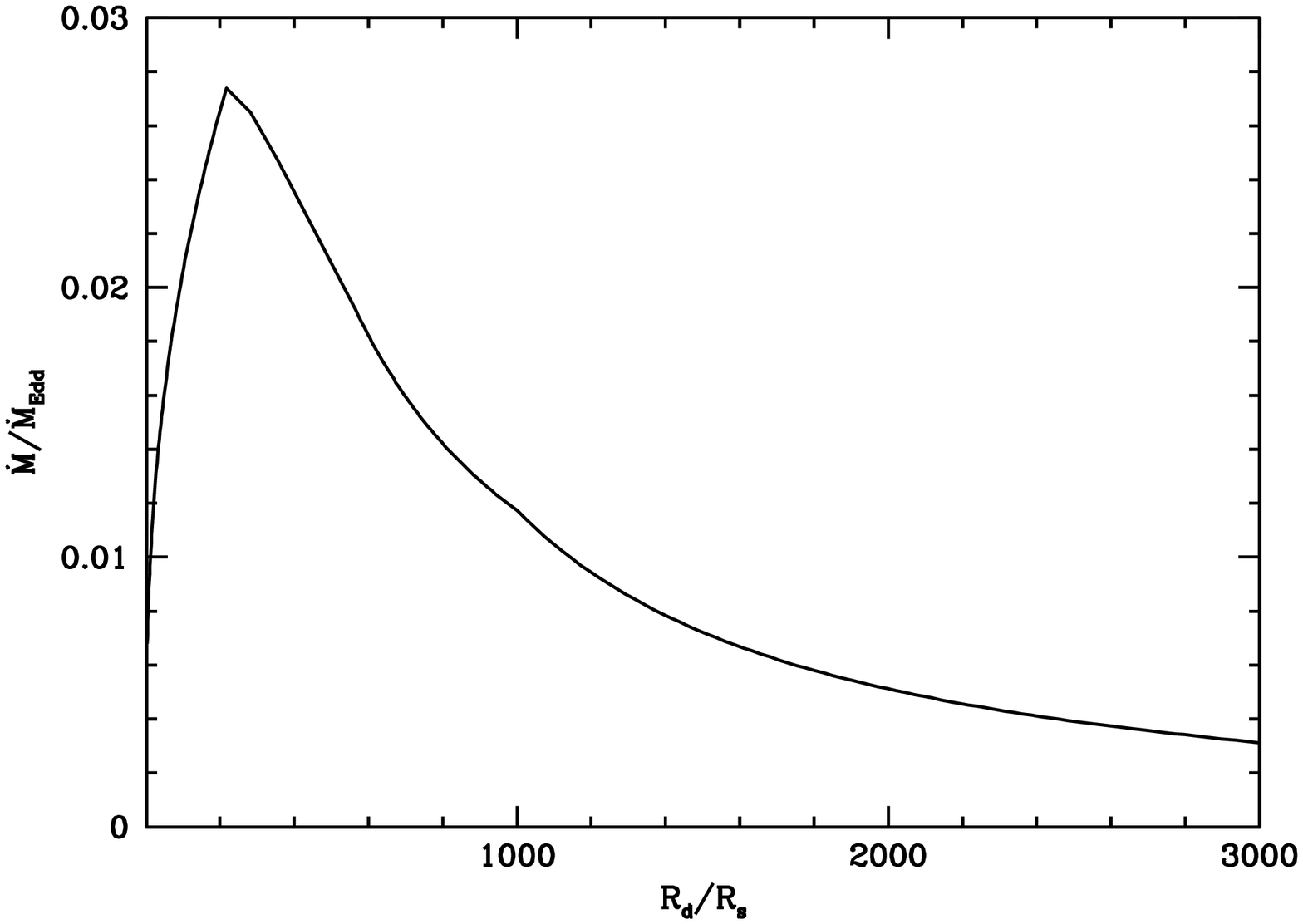}
\caption{\label{f:r_d-L}Left panel: The inner disk size, which is modified by inclusion of two cooling 
processes is shown as a function of the total luminosity. The size of the inner disk is large 
at high luminosity. At a given luminosity, the inner disk size increases with decreasing values 
of the viscosity parameter $\alpha$. Right panel: The outer extent of the inner cool disk and 
the inner extent of the outer cool disk are illustrated as a function of the mass accretion rate 
for a given viscosity parameter, $\alpha=0.2$. For a given accretion rate, the difference between these 
two radii yields the spatial extent of the coronal gap region.  The curve combines the condensation 
model described in \S 2 for the inner disk and the disk evaporation model for the inward extent 
of the outer disk based on Liu et al. (2002).} 
\end{figure}

The size of the inner disk is taken to be limited by the critical radius where neither evaporation 
nor condensation occurs. Externally, gas from the disk evaporates into the corona, while interior 
to this radius the coronal gas condenses onto the disk as a consequence of the disk corona interaction.  
For greater mass flow rates in the corona, more gas condenses onto the disk and the 
critical radius of the inner disk increases.  To illustrate this relationship, the size is 
calculated for a range of mass accretion rates.  The radial extent of the inner disk is  
modified by the inclusion of the two cooling processes since the additional cooling results in a lower 
heat flux down to the transition layer.  However, the over-cooling by bremsstrahlung radiation is 
partially compensated by the heating associated with the enthalpy flux from the condensation. 
As a consequence, the size of the inner disk is larger than the case for which one cooling 
process was considered.  In the left panel of Figure \ref{f:r_d-L}, the inner disk 
size is shown as a function of the luminosity.  It can be seen that the inner disk extends 
to greater distances with increasing luminosities or mass accretion rates.  In the right 
panel of Figure \ref{f:r_d-L}, the critical radii of the optically thick disk components are illustrated 
as a function of the mass accretion rate. For a given accretion rate, the smaller radius corresponds 
to the outer edge of the inner disk and the larger radius corresponds to the inner edge of the outer 
disk, with their difference representing the spatial extent of the coronal gap region. 
Hence, the radial extent of the coronal gap region between the inner and 
outer optically thick disk shrinks as the luminosity increases, in agreement with the presence of
a radially continuous cold disk for sufficiently high luminosities. In this case, condensation 
is no longer the only source for disk accretion, and the contribution of the disk to 
the total luminosity is much greater than for a disk fed by condensation alone.  

The size of the inner disk is a function of the viscosity parameter as also seen in Figure 
\ref{f:r_d-L} where, at a given luminosity, the disk is smaller for larger viscosity parameters.
This trend results from the (vertical) pressure and energy balance between the disk and overlying 
corona.  Specifically, for a given accretion rate, the surface density in the corona decreases as 
$\alpha$ is increased. With a lower density the heat flux to the transition layer decreases, but 
the bremsstrahlung radiation rate is reduced even more. The energy balance between cooling and heating 
results in a reduced heating rate associated with a lower enthalpy flux. The net effect results in a 
decreased condensation rate and, hence, a smaller disk size.  We note that larger viscosity 
parameters, yielding smaller disk radii, would be preferable to be consistent with our neglect 
of X-ray irradiation effects in our present level of approximation.

\begin{figure}
\plotone{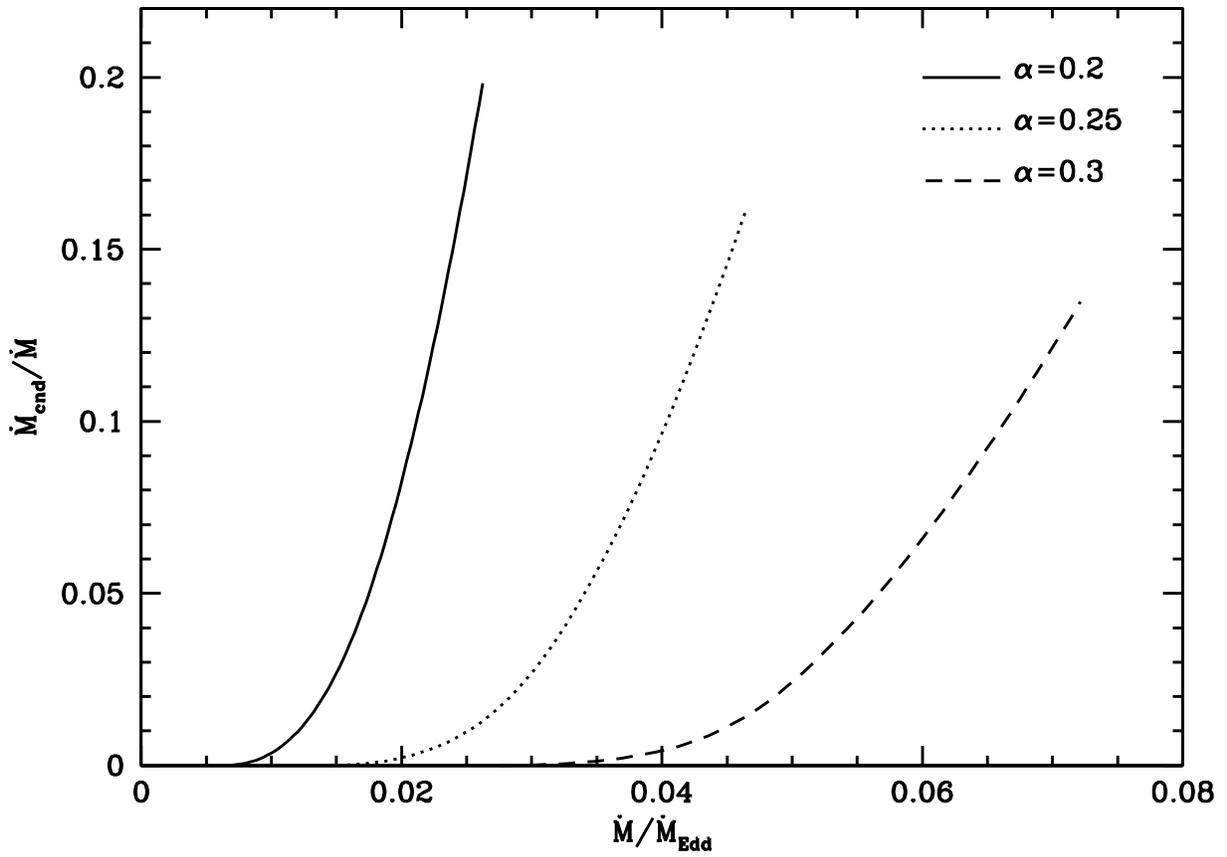}
\caption{\label{f:ratiomdot-mdot}The fraction of integrated condensation flow as a function of 
mass accretion rate for three values of the viscosity parameter $\alpha$. A greater fraction of the coronal 
flow condenses onto an inner disk at higher accretion rates,  whereas no condensation occurs at 
very low accretion rates.  This indicates a strong coronal flow can become weaker toward 
the central black hole with gas continuously condensing and accreting through a cool disk.} 
\end{figure}

The total condensation rate, which is an integral of the rate of locally condensing mass from the critical 
condensation radius to $r_{\rm tmax}=3{49\over 36}$, relative to the mass accretion rate is displayed as a function of 
the mass accretion rate in Figure \ref{f:ratiomdot-mdot}. It is evident that a large fraction 
(up to 20\%) of the coronal flow condenses onto an inner disk at high accretion rates.  As a consequence, 
the rate of mass flow in the corona decreases with decreasing distance. On the other hand, all the gas 
accretes through the corona without condensation if the mass flow rate in the corona is very low. 
The relative mass flow rate in the inner disk depends sensitively on the viscosity value.

\begin{figure}
\plotone{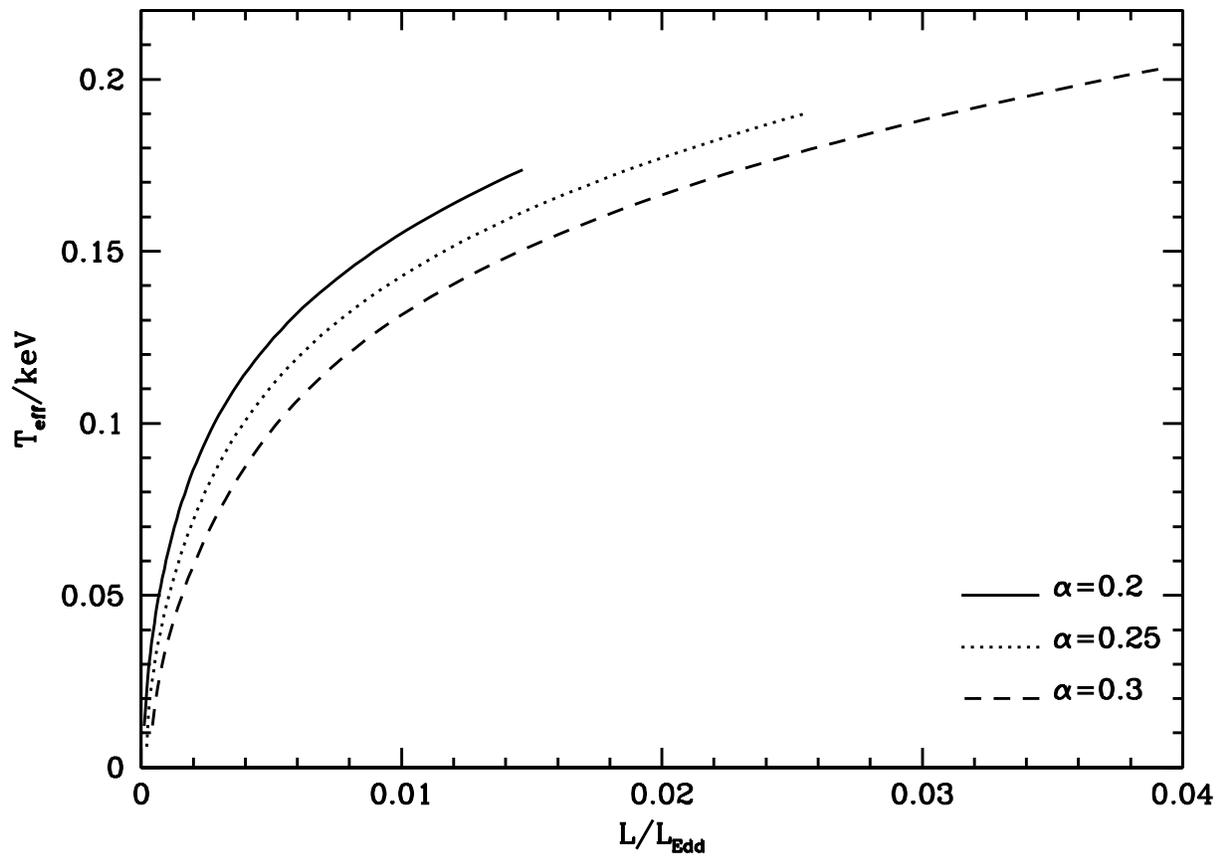}
\caption{\label{f:Teff-L}The maximal disk effective temperature as a function of luminosity 
for three values of the viscosity parameter $\alpha$.  The disk is heated by the accretion of the 
condensed gas to a high temperature at a high luminosity.  At a given luminosity, the 
temperatures are insensitive to the viscosity parameters with only a slight increase 
for lower viscosity parameters $\alpha$.} 
\end{figure}

\begin{figure}
\plotone{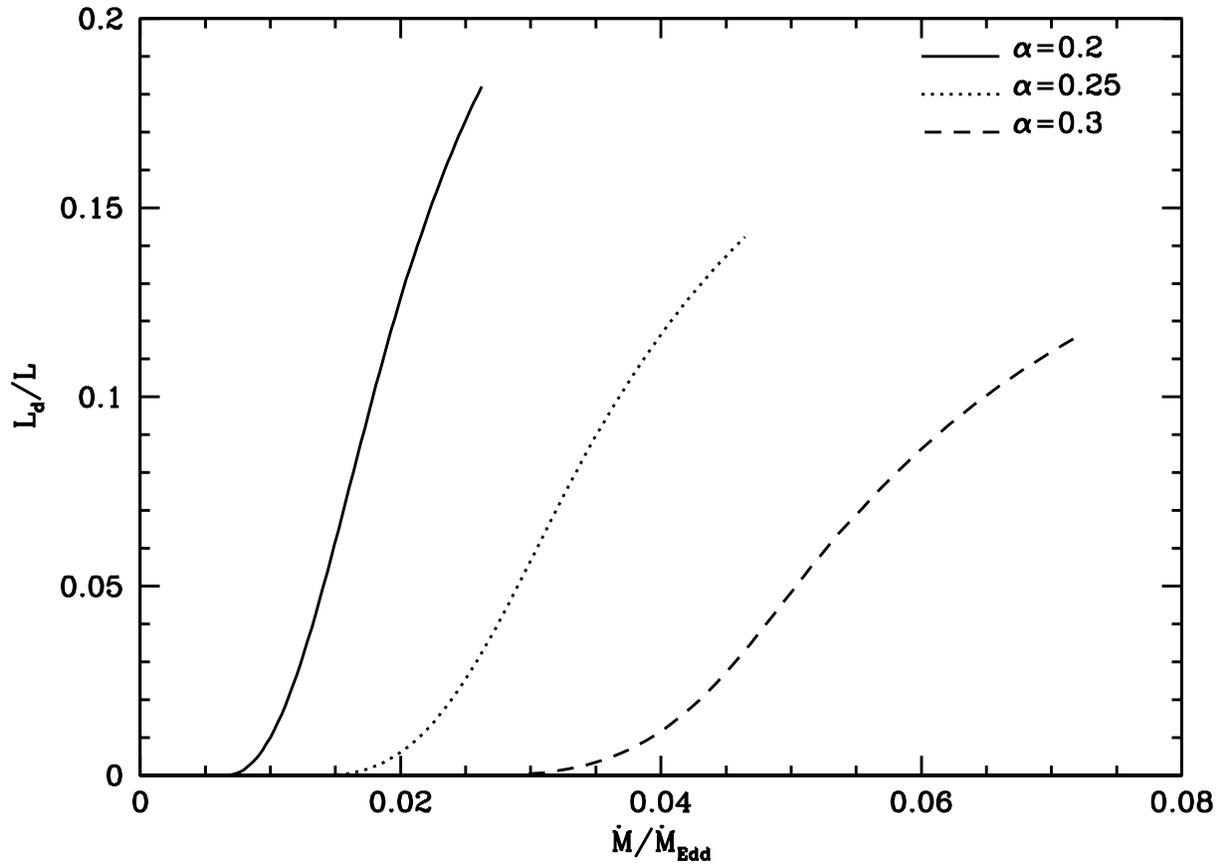}
\caption{\label{f:ratioL-mdot}The fraction of luminosity produced in the inner disk to the total 
luminosity as a function of the mass accretion rate for three values of the viscosity parameter $\alpha$. 
The contribution of the disk strengthens at higher accretion rates.}
\end{figure}

\begin{figure}
\plotone{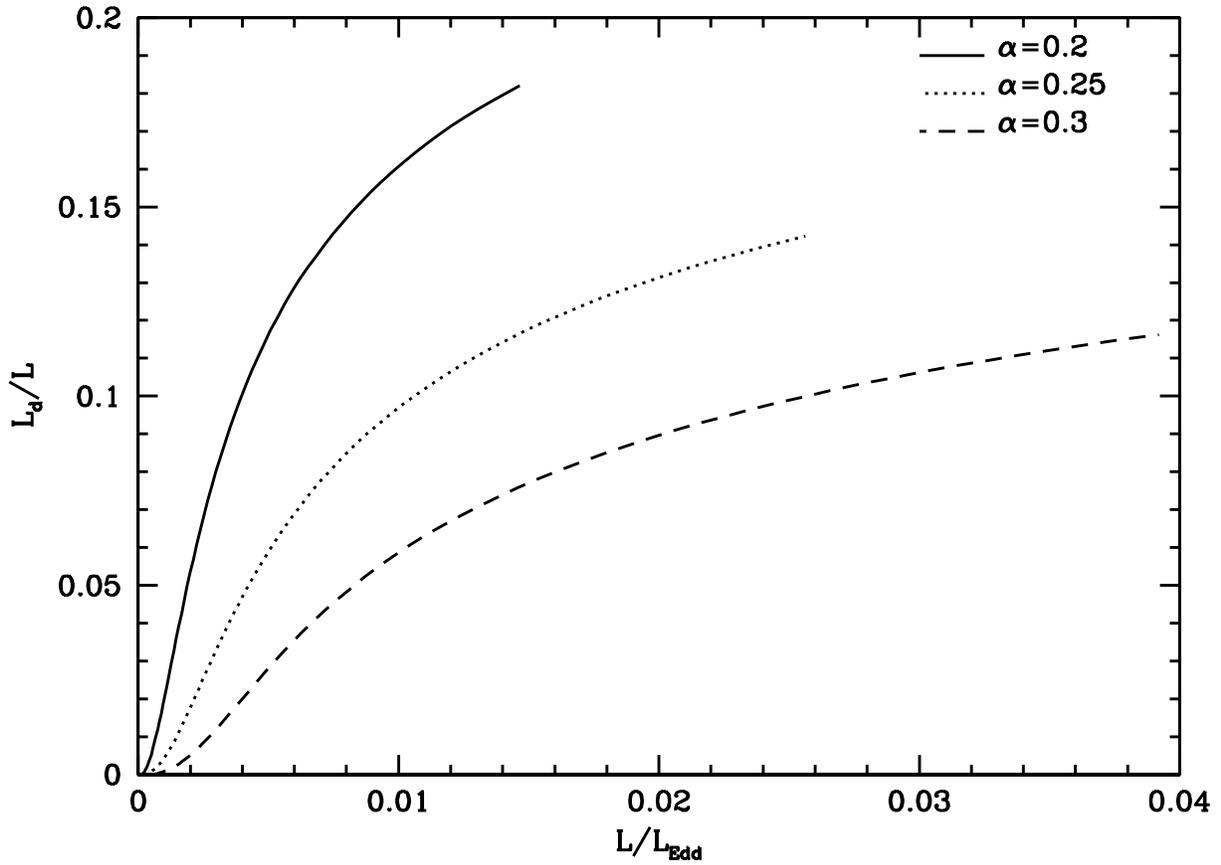}
\caption{\label{f:ratioL-L}The fraction of disk luminosity to the total luminosity as a 
function of the total luminosity for three values of the viscosity parameter $\alpha$. At low luminosities 
($L/L_{\rm Edd}\la 0.01$), the relative strength of disk component increases rapidly with the total 
luminosity. The increase is less rapid and the ratio of luminosities tends to nearly a constant value 
at higher luminosities.}
\end{figure}

The maximal disk temperature as calculated from the rate of condensation integrated over the 
entire inner disk is illustrated as a function of luminosity in Figure \ref{f:Teff-L}.  It is 
evident that the temperature increases as the total luminosity increases. Moreover, the luminosity 
from the disk component increases at higher accretion rates and luminosities as shown in Figure 
\ref{f:ratioL-mdot} and Figure \ref{f:ratioL-L}.  This feature, together with the increasing
disk size, implies that the corona/ADAF dominant flow in the low state naturally evolves to a disk 
dominant flow at high states consistent with ``standard'' picture developed for the different spectral 
states of X-ray binaries, although the exact transition is likely to be affected by the inclusion 
of X-ray irradiation of the hot coronal gas on the cool disk.  

The increasing condensation rate with accretion rate or luminosity is a direct result of the 
efficient cooling at high accretion rates. Specifically for higher accretion rates, the density 
correspondingly increases ($n \propto \dot m$), which greatly enhances the bremsstrahlung 
cooling rate ($\propto n^2$). In the transition layer, the conductive heating rate approximately 
increases with the viscous heating rate ($\propto n$).  To maintain energy balance, additional 
heating, i.e., the condensation enthalpy rate, must increase with density more rapidly than 
$\propto n^2$. As the condensation rate affects the luminosity of the inner disk, while the total 
luminosity consists of contributions from bremsstrahlung, Compton and disk radiation, the disk 
luminosity increases more rapidly than the total luminosity. Thus, their ratio increases with the 
luminosity.

The numerical results reveal that condensation occurs only for accretion rates exceeding a critical 
value, which is found to depend on the viscosity parameter.  In Figures 3-7, the effect of 
condensation was plotted from the lowest accretion rates found in the numerical calculations.
In fact, the corresponding luminosity at the critical accretion rate is from a pure ADAF rather than 
from the small disk region as shown in the figures. Therefore, the lower limit to the luminosity for 
which an inner disk could exist may well be as low as 0.001 $L_{\rm Edd}$, depending on the radiative 
efficiency of the ADAF.  

\section{Comparison to Observations}

The geometry of the accretion flow in the low hard states of X-ray transient systems has yet 
to be clarified since observational indications exist that cool material can be present in the 
inner regions where the reflection of X-rays and the formation of broad iron lines can 
contribute to the spectrum.  For the bright portion of hard-state systems near the transition 
from a spectrally hard to soft state, continuous emission and reflection components from a 
geometrically thin, optically thick inner disk are theoretically expected. Although the observational 
signatures for such a disk component 
are difficult to detect, evidence for such a feature has been inferred from the source GX 339-4 
during its 2004 outburst (Miller et al. 2006a, b) by the presence of a weak, soft, thermal component 
and by the suggestion of blurred reflection and iron lines. SWIFT J1753.5-0127 also reveals such a 
weak disk component (Miller et al. 2006a; Ramadevi \& Seetha 2007; Soleri et al. 2008), however 
iron features were not detected (Miller et al. 2006a).  For systems at very low luminosities, the 
general theoretical consensus favors a disk described by a pure ADAF solution in the inner region. 
In support of this picture, high quality spectra for several systems in quiescence at luminosities 
$L/L_{\rm Edd} \lesssim 10^{-6}$ have not revealed any evidence for a thermal component or iron 
features (e.g.  Bradley et al. 2007; Corbel et al. 2006).

Recently, Tomsick et al. (2008) have detected a weak thermal component from the source GX 339-4 
at low luminosities ($L \lesssim 0.03 L_{\rm Edd}$) based on observations they had obtained using SWIFT and 
RXTE. In addition, broad features due to iron K transitions that are likely produced by reflection 
of hard X-rays from the optically thick material near the ISCO have also been inferred from the 
spectrum. Here, we compare the results from our model predictions with these observational data. Of 
particular relevance for comparison with our theoretical analysis are the two observations of GX 339-4 
after the source underwent a transition to the hard state carried out by Tomsick et al. (2008).

Specifically, Tomsick et al. (2008) carried out two observations of GX 339-4 after the source had 
undergone a transition to the hard state.  One spectrum was obtained a few days after the transition 
(called spectrum \#1) and another spectrum was obtained 2-3 weeks later (called spectrum \#2). The 
inferred luminosity of spectrum \#1 in the energy range from 1 to 100 keV was 0.023 times the Eddington 
value based on simultaneous observations obtained with the XRT, PCA and HEXTE instruments. Spectral 
fitting indicated the presence of a cool, optically thick disk located near the ISCO with a  temperature 
of 0.201 keV. A similar analysis of spectrum \#2 yielded a luminosity of 0.008 $L_{\rm Edd}$ and a temperature 
of the inner disk of 0.165 keV. Here, for convenient comparison we adopt a source distance of 8 kpc and 
a black hole mass of $5.8 \msun$, which were used by Tomsick et al. (2008) and Miller et al. (2006b).  

In order to fit the observed luminosity and disk temperature, the accretion rate $\dot m$ and viscosity 
parameter $\alpha$ were varied until agreement between the theoretical 
modelling and the data deduced from observations was achieved. In general, the disk can be described 
in one of four different configurations depending upon the chosen values of the observables.  In particular, 
for systems characterized by very low luminosities and disk temperatures (e.g. with coronal luminosities, 
$L_{\rm c}/L_{\rm Edd}=0.001$ and $T_{\rm eff}=0.04$ keV) in the quiescent state, condensation is 
unimportant and the disk is described by an ADAF without the presence of an optically thick inner disk.  
At the other extreme, for systems at very high luminosities, say, $L_{\rm c}/L_{\rm Edd}\sim 0.1$, the 
mass accretion rate exceeds the critical rate above which an ADAF does not exist. The third configuration 
corresponds to the case where the spatial extent of the inner optically thick disk connects to the outer 
optically thick disk indicating that the disk is geometrically thin throughout and is no longer 
condensation-fed (a limiting case of configuration 4, see below). This occurs as a consequence 
of a relatively high mass accretion rate and low viscosity for systems with intermediate luminosities 
and relatively high temperatures. For instance, taking 
$L_{\rm c}/L_{\rm Edd}=0.03$ and $T_{\rm eff}=0.26$ keV, it is found that the accretion rate and viscous 
parameter are 0.035 and 0.2 respectively. However, the derived size of the disk is $417R_{\rm S}$, 
implying that the coronal gap region is absent. Only for values of $\dot m$ and $\alpha$ in the appropriate
range where the derived size of the inner disk is not very large (the fourth configuration) is the fit 
regarded as consistent with 
observations.  

\begin{figure}
\plotone{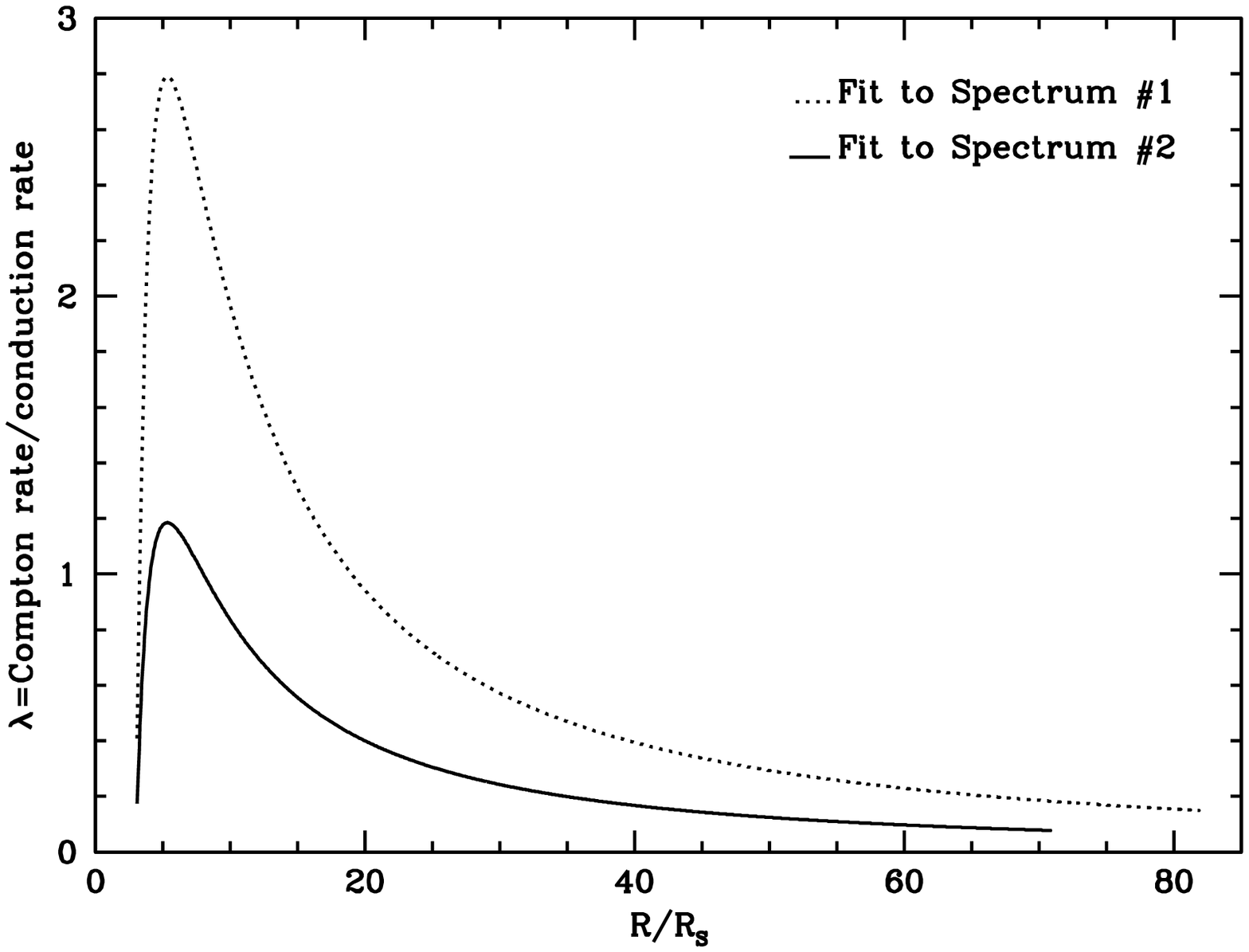}
\caption{\label{f:lambda-r}The cooling rate of the Compton process relative to that of the conduction process 
as a function of distance in the fits to spectrum \#1 and \#2. It can be seen that Compton cooling is 
stronger than conduction cooling in the model for spectrum \#1 where a disk thermal component 
of 0.2 keV is fit. In the model for spectrum \#2 the Compton cooling is comparable with conduction only 
in a small region and is much lower in the outer region.} 
\end{figure}

Our fitting procedure leads to the resulting parameters for spectrum \#1 and \#2 as listed 
in Table 1. Given $\alpha=0.31$ and $\dot m=0.0687$, we derive a coronal luminosity of 0.023 $\ledd$, a 
disk temperature of 0.203 keV, and an inner disk size of 71 $R_S$.  Similarly, given $\alpha=0.23$ and 
$\dot m=0.0299$, we derive a coronal luminosity of 0.008 $\ledd$, a disk temperature of 0.164 keV, and an 
inner disk size of 82 $R_S$. These results are in good agreement with the inferred disk parameters 
based on the two observations of Tomsick et al. (2008). With these two fits the relative strength of the 
cool disk to the hot corona can be estimated.  As listed in Table 1, the ratio of the luminosities 
associated with the thermal component, $L_{\rm d}$, to the total luminosity, $L$, is $\sim 10\%$, 
indicating a weak disk in the two observations.  In 
addition, the radii delineating the regions where Compton cooling or conduction is important are listed in 
Table 1.  For the quantitative comparison of 
cooling via Compton scattering  and thermal conduction the ratio $\lambda$ 
is shown in Figure \ref{f:lambda-r} as a function of radius for these two cases. As the radial 
distribution of $\lambda$ only depends on the condensation rate (see Eq.\ref{lambda}), i.e., the 
disk temperature listed in Table 1, the values of $\lambda$ in the two fits differ by a factor of 
$\sim 2.3$ at any given distance. Since it seems plausible to have 
the same value of $\alpha$ for both cases 
we have also explored solutions for the same value of $\alpha$ (see Table 1). Here, the observed 
luminosities are fit, but we permit small differences between the predicted and observed temperatures. It 
can be seen that the fit can still be taken as consistent with observations. The distinction between the 
fits for $\alpha = 0.25$ in the two spectra can be seen from the extent of the Compton dominated region 
as shown in Table 1. Specifically, the cooling in the corona for the fit to spectrum \#2 is dominated by 
conduction throughout the inner disk, whereas the fit to spectrum \#1 is dominated by Compton scattering 
in the inner region.  This is a consequence of the differing disk temperatures as listed in Table 1. 
This also leads to the greater sensitivity of the luminosities for a small change in accretion rates in 
the two fits. Since the disk temperature determines the energy density of the soft photon field for 
Compton scattering ($u\propto T_{\rm eff}^4$), the Compton luminosity in the fit to spectrum \#1 is much 
larger than in spectrum \#2, hence, becoming the dominant contribution to the coronal luminosity. Therefore, 
the radiation losses are more efficient when Compton cooling becomes important. 

The fits to the two observations reveal that the temperature of the inner disk is positively correlated 
with the luminosity.  We note the inferred reflection covering factors of Tomsick et al. (2008) as well 
as Miller et al. (2006b) are comparable (i.e., the covering factor is inferred as 0.22 from spectrum \#1 
and 0.24 from spectrum \#2 and 0.22 from the 2004 observations (Miller et al. 2006b)), which may suggest a comparable spatial 
extent of the inner disk.  Such an insensitivity to the luminosity could be understood if the viscosity 
parameter varied (see Table 1). 

\section{Discussion}

The structure of the inner disk region of accretion disks have been investigated for black hole 
X-ray transient systems in their low luminosity quiescent state within the context of an 
$\alpha$ viscosity model.  Particular attention has been 
focused on the conditions under which a weak, soft thermal component would be present in this state.  
By generalizing the study of Liu et al. (2007) to include the cooling associated with both conduction 
and Compton processes together, we have examined the properties of an optically thick inner disk 
fed by the condensation of gas from a corona for a range of mass accretion rates.  It has
been shown that such a disk component can be present with its spatial extent, temperature, and 
luminosity varying continuously and decreasing with a decreasing mass accretion rate.
Despite its existence and the higher conversion efficiency of gravitational energy into radiation in 
the cool disk, the contribution of the disk component to the total luminosity can be very low due to 
the low condensation rate.  Throughout the quiescent state, the mass flow rate to the central black 
hole from the corona always exceeds the mass flow rate in the inner cool disk. Hence, the disk 
structure remains geometrically thick even in the innermost regions although a soft thermal component 
is present in the spectrum.  

In our analysis, we have neglected the effect of X-ray irradiation to facilitate comparisons
with our earlier work.  Although the qualitative features of our model will not change, the parameters
describing the quantitative comparisons will be modified.  In particular, the inclusion of X-ray 
irradiation can affect the fitting parameters $\alpha$ and $\dot m$ since the observed disk 
temperature will not solely be determined by the mass flow in the inner disk. For example, for ADAF 
emission from the innermost region at height $H=R_{\rm ISCO}=3R_s$, the flux irradiating the disk 
is given by 
\begin{equation}
F_{\rm ir}(r)={{1\over 2}L_c\over 4\pi R^2}{H\over R}=1.716\times 10^{26} {\rm erg\,s^{-1}\,cm^{-2}}
m^{-1}r^{-3}{L_c\over L_{\rm Edd}}.
\end{equation}
Adopting a mean photon energy of 50keV, about one tenth of the irradiation flux (the ratio of the 
mean photon energy to electron rest mass energy) is absorbed by the disk through Compton scattering 
collisions with electrons in the disk.  Although the K-shell absorption of the metals is negligible 
at 50 keV, it can become significant at lower energies since the absorption cross section is 
proportional to $(m_ec^2/h\nu)^{7/2}$. For a photon distribution of $n_\nu\propto \nu^{-1.6}$, the 
flux absorbed by K-shell absorption is found to be $\sim 30\%$ of the total irradiation flux. Thus, 
about 40\% of the total irradiation flux is absorbed by the K-shell absorption and Compton scattering. 
The re-radiation of the absorbed flux at the inner disk surface yields an effective temperature of
\begin{equation}
T_{\rm ir}(r)=\left({0.4F_{\rm ir}\over \sigma}\right)^{1/4}=2.86keVm^{-1/4}r^{-3/4}
\left({L_c\over L_{\rm Edd}}\right)^{1/4}.
\end{equation}
At the distance of maximal temperature, $r_{tmax}=3{49\over 36}$, the effective temperature caused 
by irradiation alone for spectrum\#1 is 0.25keV and for Spectrum\#2 is 0.19keV (see \S4). These 
temperatures are slightly higher than the temperatures generated by disk accretion and could contribute 
to the observationally inferred temperature.  We note that this theoretical estimate is very 
approximate since an accurate calculation depends on the region of origin of the ADAF flux and 
the size of the coronal gap region. From Eqs.\ref{L_cmp1} and \ref{L_cmp2} it can be seen that 
the corona/ADAF radiation is dominant from a region outside the ISCO. In this case, the irradiation 
flux incident on the innermost disk decreases with distance of the source, $F_{\rm ir}
\propto r^{-2}$. For instance, assuming the ADAF radiation is dominant from $5R_s$ rather than $3R_s$, 
the irradiation flux absorbed by the disk around the ISCO is 7\% of the total flux. On the other hand, 
many coronal photons escape without interception by the disk for a large coronal gap, thereby, reducing 
the thermal effects associated with the irradiation. In the case for Spectrum \#1 and \#2 
of GX339-4, the reflection covering factor inferred from observations is $\sim 20\%$ of the half 
sky, indicating that, indeed, a very large fraction of the coronal photons escape, in contrast to 
the case of Haardt \& Maraschi (1991).  Hence, these estimates indicate that our results are 
indicative and that additional freedom exists in determining $\dot m$ and $\alpha$ when irradiation 
is taken into account.

For sufficiently low mass accretion rates, the cool inner disk vanishes.  An estimate of the mass 
accretion rate for which this occurs depends on whether condensation occurs in the vicinity of the 
ISCO. If condensation does not occur, accretion and/or evaporation can deplete the inner disk within 
a viscous time.  In the case of cooling dominated by conduction,  the critical condensation radius is 
given by $r_d=0.815\alpha^{-28/3}\dot m^{8/3}\ge 3$.  Thus, the lower limit to the accretion rate is
\begin{equation}\label{llimit1}
\dot m = 0.006 \({\alpha\over 0.2}\)^{7/2}.
\end {equation}
For an accretion rate just above this lower limit, both the size of the inner disk and the
condensation rate are very small, leading to a very low disk temperature. In this case, the Compton
cooling is negligible, and the lower limit predicted by the conduction model is self-consistent.

For disk temperatures exceeding 0.14 keV, on the other hand, Compton cooling dominates in a corona 
around a $10 M_\odot$ black hole.  If this temperature is produced by the accretion of condensed gas, 
the condensation rate would be greater than 0.002 $\dot M_{\rm Edd}$. However, this occurs at high 
accretion rates and is greater than the lower limit given by conduction model for $0.1 \le \alpha\le 
0.4$ (King et al. 2007).  Thus, the lower limit to the mass accretion rate for maintenance of an inner 
disk is set by the conduction model, i.e.  equation (\ref{llimit1}). Notwithstanding the possible 
importance of irradiation of the inner disk by the corona in providing some latitude in disk parameters, 
$\alpha$ and $\dot m$ (see above), when fitting the observational data, a determination of the mass 
flow rate below which the cool disk component vanishes could provide a constraint, within the 
framework of the condensation model, on the $\alpha$ viscosity parameter during the quiescent state. 

In the future, global calculations of the full radial and vertical disk structure should be considered
taking into account the effect of X-ray irradiation of the hot coronal gas on the inner disk and 
of the cooling associated with both the Compton and conductive processes in order to quantitatively 
confirm the picture presented in this paper.  With such models in hand, a detailed comparison of 
the spectra from such disks with observations can be undertaken. 

\acknowledgments
Financial support for this work is provided by the Theoretical Institute
for Advanced Research in Astrophysics (TIARA) operated under
Academia Sinica and the National Science Council Excellence
Projects program in Taiwan administered through grant number NSC
96-2752-M-007-007-PAE. In addition, B.F. Liu acknowledges the support by
the National Natural Science Foundation of China (grants 10533050 and 10773028).

\clearpage

\begin{table}[h]
\caption{\label{t:fits}Fitting results for GX 339-4}
\begin{tabular}{cccccccc}
\tableline\tableline
$\alpha$&$\dot m$&$L_{\rm c}/L_{\rm Edd}$&$L/L_{\rm Edd}$&$L_{\rm d}/L$(\%)&$T_{\rm eff}$(keV)&$r_d$&Compton region\\
\tableline
\multicolumn{8}{c}{Variable viscous parameter}\\
\tableline
0.31&0.0687&0.023&0.0253&9.3&0.203&71&$3.3\la r\la 19$\\
0.23&0.0299&0.008&0.009&11.2&0.164&82&$4.0\la r \la 8$\\
\tableline
\multicolumn{8}{c}{ Fixed viscous parameter}\\
\tableline
0.25&0.0471&0.023&0.0269&14.4&0.221&142&$3.2\la r\la 25$\\
0.25&0.0350&0.008&0.0088&9.0&0.156&61&no Compton region\\
\tableline
\end{tabular}
\tablecomments{Here $L_{\rm c}$, $L$,  $L_{\rm d}$, $T_{\rm eff}$, and $r_d$ are the model predictions 
for the corona luminosity, total luminosity, disk luminosity,  disk temperature and disk size for given 
viscous parameter $\alpha$ and accretion rate $\dot m$.}
\end{table} 

\clearpage

\begin{thebibliography}{}
\bibitem[]{711} Bradley, C.K., Hynes, R.I., Kong, A.K.H. et al. 2007, \apj, 667, 427
\bibitem[]{712} Corbel, S., Tomsick, J. A., \& Kaaret, P. 2006, \apj, 636, 971
\bibitem[]{713} Dullemond, C. P., \& Spruit, H. C. 2005, \aap, 434, 415
\bibitem[]{756} Esin, A.A., McClintock, J.E., \& Narayan, R. 1997, \apj, 489,865
\bibitem[]{757} Esin, A.A., Narayan, R., \& Cui, W. et al. 1998, \apj,  505, 854
\bibitem[]{758} Esin, A.A., McClintock, J.E., \& Drake, J.J. et al. 2001, \apj, 555, 483
\bibitem[]{} Haardt, F. \& Maraschi, L. 1991, \apj, 380, L51
\bibitem[]{714} Honma, F. 1996, \pasj, 48, 77
\bibitem[]{715} King, A. R., Pringle, J. E., \& Livio, M. 2007, \mnras, 376, 1740
\bibitem[]{716} Liu, B. F., Mineshige, S., \& Meyer, F., Meyer-Hofmeister,E., \& 
Kawaguchi, T. 2002, \apj, 575, 117
\bibitem[]{718} Liu, B. F., Meyer, F., \& Meyer-Hofmeister, E. 2006, \aap, 454, L9
\bibitem[]{719} Liu, B. F., Taam, R. E., Meyer, F., \& Meyer-Hofmeister, E. 2007, \apj, 671, 695
\bibitem[]{765} Maccarone, T. J. 2003, \aap, 409, 697
\bibitem[]{720} Manmoto, T., \& Kato, S. 2000, \apj, 538, 295
\bibitem[]{721} Mayer, M. \& Pringle, J. E., 2007, \mnras, 376, 435
\bibitem[]{722} McClintock, J.E., \& Remillard, R.A., 2006, In: Compact Stellar X-ray Sources, 
eds. W.H.G. Lewin, M. van der Klis, Cambridge Univ. Press
\bibitem[]{724} Meyer, F., Liu, B. F., \& Meyer-Hofmeister, E. 2000, \aap, 361, 175
\bibitem[]{725} Meyer, F., Liu, B. F., \& Meyer-Hofmeister, E. 2007, \aap, 463, 1
\bibitem[]{726} Meyer, F., \& Meyer-Hofmeister, E. 1994, \aap, 288, 175
\bibitem[]{727} Miller, J. M., Homan, J., \& Miniutti, G. 2006a, \apj, 652, L113
\bibitem[]{728} Miller, J. M., Homan, J., Steeghs, D., Rupen, M., Hunstead,
R. W., Wijnands, R., Charles, P. A., \& Fabian, A. C. 2006b, \apj, 653, 525
\bibitem[]{730} Narayan, R., \& Yi, I. 1994, \apj, 428, L13
\bibitem[]{731} Narayan, R., \& Yi, I. 1995a, \apj, 444, 231
\bibitem[]{732} Narayan, R., \& Yi, I. 1995b, \apj, 452, 710
\bibitem[]{733} Ramadevi, M. C., \& Seetha, S. 2007, \mnras, 378, 182
\bibitem[]{734} R\`o\.za\`nska, A., \& Czerny, B. 2000, \aap, 360, 1170
\bibitem[]{735} Rykoff, E. S., Miller, J. M., Steeghs, D., \& Torres, M. A. P. 2007, \apj, 666, 1129
\bibitem[]{736} Shakura, N. I., \& Sunyaev, R. A. 1973, \aap, 24, 337
\bibitem[]{737} Soleri, P., Altamirano, D., Fender, R. et al. 2008, astro-ph/0803.1735 
\bibitem[]{738} Spitzer L., 1962, Physics of Fully Ionized Gases,
2nd edition, Interscience Publ., New York, London
\bibitem[]{740} Spruit, H. C., \& Deufel, B. 2002, \aap, 387, 918
\bibitem[]{741} Stepney, S. 1983, \mnras, 202, 467
\bibitem[]{742} Sutherland, R.S. \& Dopita, M.A. 1993,\apjs, 88, 253
\bibitem[]{743} Tomsick, J. A. et al. 2008, \apj, in press (astro-ph/0802.3357)
\end{thebibliography}
\end{document}